\documentclass[prb,aps,amsmath,amssymb,amsfonts,floatfix,groupeaddress,showpacs,twocolumn]{revtex4-1}
\usepackage{longtable}
\usepackage{graphicx}
\usepackage{epsfig}
\usepackage[dvips]{color}
\usepackage{dcolumn}
\usepackage{nicefrac}

\newcommand{\iomn}{i\omega_n}

\newcommand{\uR}{\underline{R}}
\newcommand{\uk}{\underline{k}}
\newcommand{\ur}{\underline{r}}
\newcommand{\bP}{\bar{P}}

\def\beq{\begin{equation}}
\def\eeq{\end{equation}}
\def\be{\begin{equation}}
\def\ee{\end{equation}}
\def\iomn{i\omega_n}

\def\t{\mbox{tr}\,}
\def\cG0{{\cal G}_0}
\def\cG{{\cal G}}

\def\a{\alpha}

\def\uc2{$U_{c2}$}
\def\uc1{$U_{c1}$}
\def\bavs3{BaVS$_3$}

\def\t2g{$t_{2g}$}

\def\a1g{$a_{1g}$}

\begin{document}

\title{Approaching finite-temperature phase diagrams of strongly
correlated materials:\\
a case study for V$_2$O$_3$}

\author{Daniel~Grieger, Christoph~Piefke, Oleg~E.~Peil and Frank~Lechermann}

\affiliation{I. Institut f\"ur Theoretische Physik,
  Universit\"at~Hamburg, Jungiusstr.~9, D-20355~Hamburg, Germany}

\begin{abstract}
Examining phase stabilities and phase equilibria in strongly correlated materials asks
for a next level in the many-body extensions to the local-density 
approximation (LDA) beyond mainly spectroscopic assessments. Here we put the
charge self-consistent LDA+dynamical mean-field theory (DMFT) methodology based
on projected local orbitals for the LDA+DMFT interface and a tailored pseudopotential
framework into action in order to address such thermodynamics of realistic strongly 
correlated systems. Namely a case study for the electronic phase diagram of the
well-known prototype Mott-phenomena system V$_2$O$_3$ at higher temperatures is
presented. We are able to describe the first-order metal-to-insulator transitions 
with negative pressure and temperature from the self-consistent computation
of the correlated total energy in line with experimental findings. 
\end{abstract}
\pacs{71.45.Gm, 71.45.Lr, 71.30.+h, 73.20.Mf, 71.15.Mb}
\maketitle

\section{Introduction}
The first-principles computation of phase diagrams at finite temperature $T$
for multi-component materials systems is a quite formidable challenge. 
Although there are very successful (semi-)empirical methodologies to 
compute the thermodynamics of binary (or higher) realistic systems, most 
notably the CALPHAD approach,~\cite{sau98} the capability of predicting phase 
diagrams by starting from an ab-initio quantum-mechanical level has a rather 
strong appeal to many theorists. 
Dating back to the pioneering work in this research area by 
Hume-Rothery in the 1930's through empirical rules based on atomic sizes and 
electronegativities,~\cite{hum35,hum69} the field has reached quite a level 
of sophistication. After the extension of Hume-Rothery's original ideas by 
Miedema and coworkers~\cite{mie73} via additionally introducing the 
electronic charge density in the determination of the formation energy, 
approaches build on density functional theory (DFT) in the Kohn-Sham 
representation~\cite{koh65} eventually have taken over and have been dominating the research on 
atomistic phase-diagram calculations~\cite{wil83,gyo83,wei87,car88} since the mid 
1980s (see e.g. Ref.~\onlinecite{mue03} for a review).

However, materials systems with less-screened Coulomb interactions among the 
electrons of the order of or larger than the bandwidth $W$ have remained so far out 
of reach. Conventional representations of DFT, via e.g. the local-density 
approximation (LDA),~\cite{koh65} are not capable of accounting for the effects 
of strong electronic correlations. Phase transformations at finite $T$ either of 
pure electronic kind or driven by electronic correlations are usually not
describable solely within Bloch band theory. On the other hand, many novel materials
which are technologically promising, because of e.g. enhanced response behavior,
display signatures of strong correlations. Furthermore, even well-known allotropes
of transition metals or prominent transition-metal alloys with or close to
magnetic order (like e.g. the iron-aluminum system~\cite{lec02}) are rather difficult 
to model within standard LDA(-like) approaches due to the lack of explicit 
many-body correlation effects. 

The combination of LDA with the dynamical mean-field theory (DMFT), the
so-called LDA+DMFT approach, nowadays prosperously allows to include the effects 
of strong Coulomb interactions in realistic solids (see e.g. 
Ref.~\onlinecite{kotliar_review} for a review). Note that one may easily use
the generalized-gradient approximation (GGA) for the DFT part, but
as far as it concerns explicit strong correlation effects the difference to the LDA 
approach are usually negligible. Yet there are only few implementations 
that handle the LDA+DMFT formalism in a charge self-consistent framework, i.e. 
accounting for the feedback of the local electronic self-energy onto the charge 
density that determines the Kohn-Sham effective potential, until self-consistency 
of the complete interacting charge density is 
achieved.~\cite{kotliar_review,min05,pou07,hau10,aic11,ama12,gra11} Additionally,
in order to account for competing strongly correlated phases at elevated 
temperatures there are high demands on the accuracy and generality of the
underlying band-structure methodology as well as the utilized DMFT impurity 
solver. 
\begin{figure}[t]
\centering
\includegraphics*[width=8.5cm]{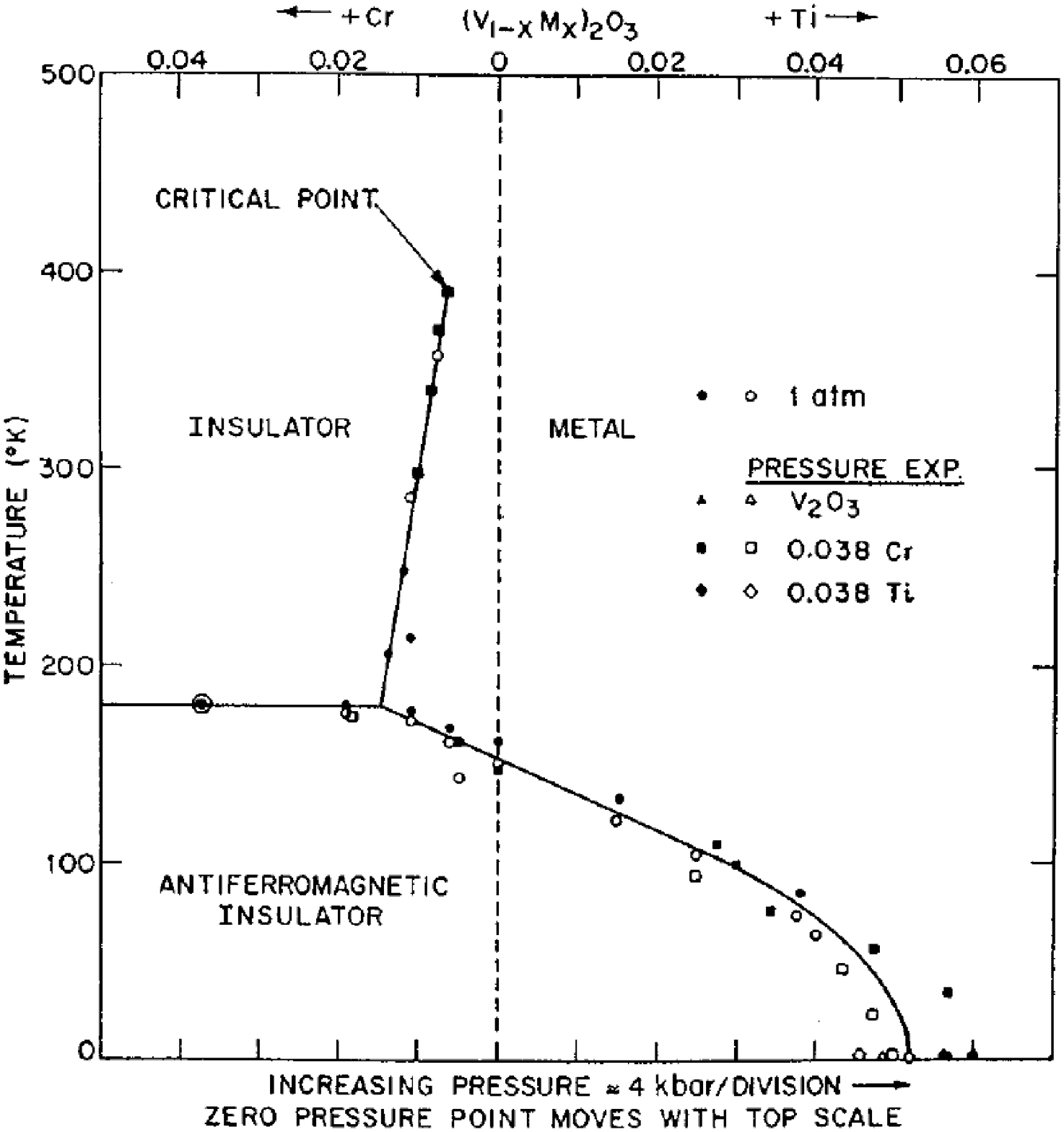}
\caption{Experimental phase diagram of V$_2$O$_3$ from
Refs.~\onlinecite{mcw71,mcw73}.
\label{fig:v2o3}}
\end{figure}

As to explicit realistic phase-competition studies, there is prominent work
within that scope mainly in the area of $f$-electron 
compounds.~\cite{kotliar_review,sav06} In a recent LDA+DMFT study, Leonov 
{\sl et al.} provided a quite successful modeling of the high-temperature 
bcc-to-fcc transition in iron~\cite{leo11} using the powerful quantum Monte Carlo method 
for the impurity solution. However, there the charge selfconsistency was
neglected and total energies have been calculated in a post-processing 
scheme. 

In the present work we want to review the current state of the art formalism for 
handling the charge self-consistent LDA+DMFT method with the direct calculation of the 
correlated total energy. The full approach is applied to the key
features of the electronic phase diagram of the famous sesquioxide V$_2$O$_3$ 
(see Fig.~\ref{fig:v2o3}) above room temperature. When revealing the phase 
boundaries between metallic and insulating phases with $T$ and pressure $p$, we 
neglect the effect of chemical disorder as well as explicit 
electronic entropy contributions. The former is not expected to play a vital role in 
the present study since we are interested in the stoichiometric system well below
possible disordering/melting temperatures. Extending the 
strongly correlated formalism to composition-dependent phase diagrams is in 
principle possible, e.g. via the so-called stat-DMFT method.~\cite{dob97} As it
will be shown, the present results in the more restricted scheme are already 
encouraging and it should be mainly a matter of time when such new and further
elaborate techniques will become a generic tool in the context of first-principles 
phase-diagram evaluations.

\section{Theoretical framework}
The charge-self-consistent scheme of the LDA+DMFT framework is technically and 
computationally rather demanding and only few implementations thereof exist up to 
now.~\cite{pou07,hau10,aic11,ama12,gra11} For the DFT(LDA) part of the calculations, 
we here employ a mixed-basis pseudopotential~\cite{mbpp_code} (MBPP) as well as a 
projector-augmented wave~\cite{blo94} (PAW) implementation. 

The explicit many-body effects are treated within a multi-orbital Hubbard model 
including also all non density-density local interaction terms, i.e. we use the
complete Coulomb matrix according to full rotational symmetry. This amounts to a 
parametrization with two interaction integrals, namely the Hubbard $U$ and the 
Hund's exchange $J$. The interacting problem is solved within DMFT using a
continuous-time quantum Monte-Carlo (CT-QMC) impurity solver in the
hybridization-expansion formulation~\cite{wer06} as implemented in the
TRIQS package.~\cite{triqs_code} 

In short, the charge-self-consistency condition 
can on equal footing be understood as a way of improving the density functional 
of DFT to include more explicit correlation effects as well as finding a 
realistic and consistent effective single-particle part of DMFT via combining 
both formalisms to a new, complete cycle, which is summarized in the following 
sections. 

Before that, however, it is vital to note that on general grounds this
LDA+DMFT formalism is manifestly temperature dependent and therefore in
principle ideally suited for thermodynamic problems. This is in stark contrast
to the standard extension of the Kohn-Sham formalism towards finite $T$ via 
the Mermin theory of including the proper Fermi function for the electronic states.~\cite{mer65} Here the full impact
of temperature on the many-body level, including e.g. the effective disappearance
of Bloch-like quasiparticle states and thus the localization due to the loss of 
coherency at large $T$, is properly taken care of.

\subsection{Projected local orbitals}
The DFT(LDA) method utilizes an orbital-independent representation of the
effective-single particle Hamiltonian for the electronic structure resulting
in Bloch-Kohn-Sham (KS) wave functions for the solid-state electronic structure.
On the other hand the DMFT equations make use of a local correlated subspace 
in order to include the effects of strong Coulomb correlations in condensed
matter. Thus the first step that has to be peformed when interfacing the DFT(LDA) 
and the DMFT technique is the extraction of a suitable correlated subspace
$\mathcal C$ starting from the complete Hilbert space of Bloch KS band states. 
This is done in the projected-local-orbitals (PLO) 
scheme.~\cite{ku02,ani05,ama08,kar11} For completeness we here briefly review the 
methodology in the context of charge-self-consistency and the total-energy
calculation. More details may be found in Ref.~\onlinecite{ama08}. 

Normalized orthogonal projections onto chosen local orbitals with character $m$
and centered at site $\uR$ may be defined via
\begin{equation}
\label{Eqn:PROJDEFINITION}
\bP^{\uR}_{m \nu} 
(\underline k) \equiv \sum_{\uR' m'} \left \lbrace \left [ \underline
{\underline O} (\underline k) \right ]^{- \frac 12} \right\}_{m m'}^{\uR\uR'} 
\,\langle\chi_{\uR m'}^{\hfill}|\Psi_{\uk\nu}^{\hfill}\rangle\;,
\end{equation}
where the $\vert \Psi_{\underline k \nu} \rangle$ for wave vector
$\uk$ and band $\nu$ are chosen to be a subset $\mathcal W$ of the
Bloch states of the original LDA treatment and $\underline {\underline
  O}$ describes the overlap matrix written as
\begin{equation}
O_{mm'}^{\uR\uR'}(\underline k) \equiv \sum_{\nu \in \mathcal W}
\langle\chi_{\uR m}^{\hfill}|\Psi_{\uk\nu}^{\hfill}\rangle
\langle\Psi_{\uk\nu}^{\hfill}|\chi_{\uR' m'}^{\hfill}\rangle\;.
\end{equation}
The set of states $\{\vert \chi_{\uR m}^{\hfill}\rangle\}$ together with the
energy window ${\cal W}$ define the correlated subspace $\mathcal C$, chosen such 
that the problem of strong local Coulomb interactions is adequately represented. 
Conveniently, $\mathcal C$ is adapted to an available localized basis used in a 
given band-structure code, i.~e. linear combinations of the mixed basis within the 
MBPP framework or the partial waves of PAW. 

Using the projections (\ref{Eqn:PROJDEFINITION}), one can construct the 
one-particle Green's function within the truncated Bloch space $\mathcal W$ 
via the double-counting corrected local DMFT self-energy (written in Matsubara 
frequencies $\omega_n$=$(2n+1)\pi k_{\mathrm B}T$), which is assumed block diagonal in the
correlated sites, written as
\begin{equation}
\Delta\underline {\underline \Sigma}^{\uR\uR'}(i\omega_n)\equiv 
\Bigl(\underline{\underline \Sigma}^{\mathrm{imp},\uR}(i \omega_n)
-\underline {\underline \Sigma}^{\mathrm{dc},\uR}\Bigr)\delta_{\uR\uR'}\;.
\end{equation}
This Bloch Green's function is thereby connected to the correlated subspace 
through an upfolding procedure, i.e.,
\begin{eqnarray}
\label{Eqn:GBLDEFINITION}
\underline {\underline G}^{\mathrm{bl}} (i \omega_n, \underline k)&=&
\Bigl[ (i \omega_n + \mu)
\underline {\underline 1} - \underline {\underline \epsilon}_{\underline k}
^{\mathrm{KS}}\nonumber\\
&&\hspace*{-0.75cm}-\,\sum_{\uR\uR'}\underline {\underline \bP}^{\uR\dagger}
(\underline k) \cdot\Delta \underline {\underline \Sigma}^{\uR\uR'}(\iomn)\cdot
\underline {\underline \bP}^{\uR'}(\underline k)\Bigr]^{-1}\;.
\end{eqnarray}
In this equation, $\underline {\underline \epsilon}_{\underline k}
^{\mathrm{KS}}$ denotes the diagonal matrix of Kohn-Sham eigenvalues for the Bloch
states and $\mu$ marks the chemical potential (see section
\ref{Sec:CHEMICALPOTENTIAL} for details). This Green's function can then be
downfolded to the correlated subspace, enforcing the DMFT self-consistency
condition (proper normalization of the $k$-sum here and in the following is 
understood)
\begin{equation}
\underline {\underline G}^{\mathrm{imp},\uR} (i \omega_n) \equiv
\sum \limits_{\underline k} \underline {\underline \bP}^{\uR} (\underline k)
\cdot \underline {\underline G}^{\mathrm{bl}} (i \omega_n, \underline k)
\cdot \underline {\underline \bP}^{\uR\dagger}(\underline k)\;.
\end{equation}
This impurity Green's function can then be used to supply a new DMFT bath 
Green's function
\begin{equation}
(\underline{\underline{\mathcal G_0}}^{\uR})^{-1}(\iomn)=
(\underline{\underline G}^{\mathrm {imp},\uR})^{-1}(\iomn)+
\underline{\underline \Sigma}^{\mathrm{imp},\uR}(\iomn)
\end{equation}
that enters the impurity solver yielding eventually an updated impurity 
self-energy until convergence is achieved. The outlined iterative scheme marks 
the usual DMFT cycle without charge-self-consistency, since it then works as 
post-processing scheme to a once computed set of Kohn-Sham objects 
$\{\epsilon^{\rm KS}_{\uk\nu},\Psi_{\uk\nu}^{\hfill}\}$.

\subsection{Expressing charge densities}
The fundamental step in the {\sl self-consistent} combination of DFT(LDA) and DMFT is
provided by the expression of the basic quantities of each method in terms of the
basic quantities of the other method. Namely, charge density for DFT and 
one-particle Green's function for DMFT. For this purpose, we define a Kohn-Sham 
Green's function through
\begin{equation}
\label{Eqn:GKSDEFINITION}
\underline {\underline G}^{\mathrm{KS}} (i \omega_n, \underline k) =
\left [ (i \omega_n + \mu_{\mathrm{KS}})
\underline {\underline 1} - \underline {\underline \epsilon}_{\underline k}
^{\mathrm{KS}}\right ] ^{-1}\;.
\end{equation}
Note that this function is in general different from the DMFT ``Weiss-Field'' 
$\mathcal G_0$. The choice of $\mu_{\mathrm{KS}}$ is described in detail in section
\ref{Sec:CHEMICALPOTENTIAL}. In the following the band indices $\nu\nu'$ live
in the truncated Bloch Hilbert space ${\cal W}$ and we drop for convenience the site
index $\uR$. Generalization of the formulae including the latter is 
straightforward and since the charge density is additive, contributions from
(supposingly weakly correlated) bands outside ${\cal W}$ are most easily taken
care of.
The trace of $\underline {\underline G}^{\mathrm{KS}} (i \omega_n, \underline k)$ 
expressed in the Bloch basis is nothing else than the charge density of a 
standalone KS-LDA calculation which reads
\begin{equation}
\rho^{\mathrm{KS}}(\underline r) = \frac 1 \beta
\sum \limits_{\underline k n \nu \nu'}
\langle \underline r \vert \Psi_{\underline k \nu} \rangle \,
G^{\mathrm {KS}}_{\nu \nu'} (i \omega_n, \underline k) \,
\langle \Psi_{\underline k \nu'} \vert \underline r \rangle\;,
\end{equation}
with $\beta$=1/$k_{\mathrm B}T$ as the inverse temperature. A very similar form can be found for 
the charge density from a post-processing DMFT calculation, i.e.
\begin{equation}
\rho(\underline r) = \frac 1 \beta
\sum \limits_{\underline k n \nu \nu'}
\langle \underline r \vert \Psi_{\underline k \nu} \rangle \,
G^{\mathrm {bl}}_{\nu \nu'} (i \omega_n, \underline k) \,
\langle \Psi_{\underline k \nu'} \vert \underline r \rangle\;.
\end{equation}
Thus the difference $\rho'(\underline r) =
\rho(\underline r) - \rho^{\mathrm{KS}}(\underline
r)$ is given by
\begin{eqnarray}
\rho'(\underline r)&=&\frac{1}{\beta}\sum \limits_{\uk n\nu\nu'}
\langle \underline r \vert \Psi_{\underline k \nu} \rangle
\Bigl\{ \underline {\underline G}^{\mathrm {KS}}(\iomn,\uk)\cdot
\Bigl( (\underline {\underline G}^{\mathrm {KS}})^{-1}(\iomn,\uk)\nonumber\\
&&\hspace*{0.5cm}
-\,(\underline {\underline G}^{\mathrm {bl}})^{-1}(\iomn,\uk) \Bigr)\cdot
\underline {\underline G}^{\mathrm {bl}}(\iomn,\uk)\Bigr\}_{\nu \nu'}
\langle \Psi_{\underline k \nu'} \vert \underline r \rangle\nonumber\\
&\equiv& \sum \limits_{\uk\nu\nu'}
\langle \underline r \vert \Psi_{\underline k \nu} \rangle\,
\Delta N_{\nu \nu'}(\underline k)\,
\langle \Psi_{\underline k \nu'} \vert \underline r \rangle\;.
\end{eqnarray}
As described in Ref.~\onlinecite{lec06}, the object $\underline
{\underline {\Delta N}}(\underline k)$ can be rewritten as
\begin{eqnarray}
\label{Eqn:DELTANDEFINITION}
\underline {\underline {\Delta N}}(\underline k)
&=& \frac 1 \beta \sum \limits_n
\Bigl\{ \underline {\underline G}^{\mathrm {KS}}(\iomn,\uk)\cdot 
\Bigl(\underline {\underline \bP}^\dagger(\underline k) \cdot
\Delta \underline {\underline \Sigma}(\iomn)\cdot
\underline {\underline \bP}(\underline k)\nonumber\\
&&\hspace*{1.25cm}-\,(\mu- \mu_{\mathrm{KS}})
\underline {\underline 1} \Bigr) \cdot
\underline {\underline G}^{\mathrm {bl}}(\iomn,\uk)\Bigr\}\;.
\end{eqnarray}
Therewith a simple representation of the total charge density including 
self-energy effects beyond LDA is provided, reading
\begin{equation}
\label{Eqn:RHOREPRESENTATION}
\rho(\underline r)=\sum \limits_{\underline k\nu\nu'}
\langle \underline r \vert \Psi_{\underline k \nu} \rangle
\Bigl(f(\tilde{\epsilon}_{\underline k \nu})\delta_{\nu \nu'}+ 
\Delta N_{\nu \nu'}(\underline k)\Bigr) 
\langle \Psi_{\underline k \nu'} \vert \underline r \rangle
\end{equation}
and to be used and manipulated in a given DFT-based band-structure code. 
Here $f(\epsilon)$ denotes
the Fermi-distribution function and $\tilde{\epsilon}_{\underline k \nu}$=$\epsilon_{\underline k \nu}$$-$$\mu_{\rm KS}$. Hence the inclusion of the DMFT self-energy
renders it necessary to not only incorporate modifying terms diagonal in the
Bloch states, but also off-diagonal contributions. The problem of truncating the 
whole Bloch space to a subspace $\mathcal W$ therefore reduces to taking into 
account the correct set of bands in each summand. Details on the implementation 
thereof in the different KS basis sets are given in appendix~\ref{App:CDDFTBASIS}.

\subsection{Self-consistency condition}
The aim of charge-self-consistency is to include DMFT self-energy
effects in the charge density, so that $\rho^{\mathrm{KS}}(\underline
r)$ and $\rho(\underline r)$ can obviously not be the same quantity.
Instead, it is most instructive to use the (spectral density-)
functional approach by Savrasov and Kotliar,~\cite{sav04}
incorporating both one-particle Greens's function and charge
density. Extremization thereof with respect to the charge density
$\rho(\underline r)$ basically yields the Kohn-Sham equations in which
the correlated charge density $\rho(\underline r)$ is used as an input
for the effective potential $\hat V_{\mathrm {eff}} \left [ \rho
  (\underline r) \right ]$, i.e.
\begin{equation}
\left [ \hat T + \hat V_{\mathrm {eff}} \left [ \rho (\underline r) \right ] -
\epsilon_{\underline k \nu} \right ] \vert \Psi_{\underline k \nu} \rangle
= 0\;.\label{Eqn:corrKS}
\end{equation}
Similarly, extremization with respect to the Green's function yields
the usual expression for the DMFT self-energy. Thus the complete cycle
can be constructed as follows: From an initial DFT(LDA) calculation,
perform conventional DMFT steps and compute a correlated charge
density $\rho(\underline r)$ as given by
eq.~(\ref{Eqn:RHOREPRESENTATION}). That charge density is reinserted
into the band-structure code (for this step, knowing the elements of
$\underline {\underline {\Delta N}}(\underline k)$ is sufficient) and
new effective single-particle wave functions $\vert \Psi_{\underline k
  \nu} \rangle$ are computed using eq.~(\ref{Eqn:corrKS}). These
finally enter eq.~(\ref{Eqn:PROJDEFINITION}) to build a new correlated
subspace for DMFT (keeping the set $\{\vert \chi_{\uR
  m}^{\hfill}\rangle\}$ unaltered). This enlarged cycle is iterated
until full charge-self-consistency is reached, i.~e. charge density
and self-energies (and thus the matrix $\underline {\underline {\Delta
    N}}^{(\underline k)}$) remain constant with iterations. Our
experience with the cycle shows that convergence is robust, a
linear mixing scheme is often sufficient.

\subsection{Chemical potential}
\label{Sec:CHEMICALPOTENTIAL}
As usual, the chemical potential $\mu$ is adjusted such that the
resulting total charge density $\rho_{\rm tot}(\underline r)$ holds the correct 
total number of electrons $N_{\rm tot}$ and is enforced via
\begin{equation}
\label{Eqn:INTRHOISN}
\int \hspace*{-0.1cm}d\underline r \,\,\rho_{\rm tot} (\underline r) 
=N_{\nu\notin {\cal W}}+\frac 1 \beta
\sum \limits_{\underline k n \nu}
G^{\mathrm {bl}}_{\nu \nu} (i \omega_n, \underline k) = N_{\rm tot}\;.
\end{equation}
In Ref.~\onlinecite{lec06} it is argued that this charge neutrality condition is
imposed on $\rho_{\rm tot}(\underline r)$ only (not on
$\rho^{\mathrm{KS}}(\underline r)$). Thus the parameter $\mu_{\mathrm{KS}}$, 
which is defined to be the energy up to which the Kohn-Sham states of the 
DFT(LDA) part are filled (with the proper Fermi-distribution function), can be 
chosen to be equal to the chemical potential $\mu$, which would mean that the 
integrated charge density changes due to correlations in 
eq.~(\ref{Eqn:RHOREPRESENTATION}) vanishes.

However, in order to clarify relations between some of the quantities
that occur in the formalism, it can be useful to choose
$\mu_{\mathrm{KS}}$ such that the DFT(LDA) part of the calculation is
already charge neutral. It can easily be shown that with the
correction term of eq.~(\ref{Eqn:DELTANDEFINITION}), this choice
does not affect $\rho(\underline r)$. Anyway, it is important to note
that $\mu_{\mathrm{KS}}$ itself has no physical interpretation in the enlarged
LDA+DMFT framework. 

\subsection{Total energy}
In order to obtain total energies within the charge-self-consistent formalism, 
the spectral density functional approach is applied again. Based thereon, the total
energy may be computed from~\cite{pou07}
\begin{equation}
E_{\mathrm{LDA+DMFT}} = E_{\mathrm{LDA}}+ \sum
\limits_{\underline k \nu} \epsilon_{\underline k \nu}^{\mathrm{KS}}
\, \Delta N_{\nu \nu} ^{(\underline k)} + \langle \hat H_U \rangle -
E_{\mathrm{dc}}\;.
\end{equation}
Note that this expression relies on the fact that the adapted diagonal
basis for the Kohn-Sham single-particle Hamiltonian is utilized, which
yields eigenvalues $\epsilon_{\underline k
  \nu}^{\mathrm{KS}}$. Several approaches are possible to obtain the
expectation value of the two-particle Hamiltonian $\langle \hat H_U
\rangle$. Here we choose to apply the Galitskii-Migdal
formula.~\cite{gal58} As shown by Boehnke {\sl et al.}~\cite{boe11},
the quality of the numerical data can be improved by choosing a
suitable basis set, i.~e. Legendre polynomials, for the represenation
of the one-particle Green's function. Note that although this LDA+DMFT
total energy is temperature-dependent, we here keep the 'energy'
notion, since for a well-defined 'free energy' a clear definition of
an entropic part would be in order.

\section{The V$_2$O$_3$ system}
\begin{figure}[t]
\centering
\includegraphics*[width=8.5cm]{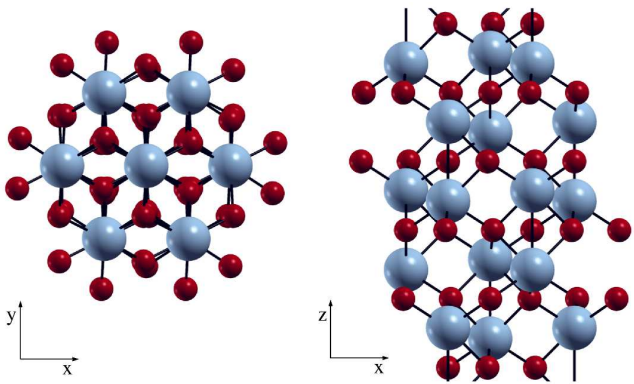}
\caption{(Color online) Corundum structure of V$_2$O$_3$ viewed along the
$z$-axis (left) and along the $y$-axis (right), with depicted V ions 
(large grey) and O ions (small red/dark).
\label{fig:struc}}
\end{figure}

The vanadium sesquioxide V$_2$O$_3$ belongs to the most prominent strongly 
correlated compounds and has already been subject to many theoretical 
efforts.~\cite{cas78,mat94,ezh99,elf03,eye05,hel01,kel04,pot07} 
At elevated temperatures it orders in the corundum structure in which there are 
V-V pairs along the crystallographic $c$-axis and a honeycomb lattice appears in 
the $xy$-plane (see Fig.~\ref{fig:struc}). The
V ions reside within an octahedron of oxygen ions, respectively, building up
a trigonal crystal field for the transition-metal ion. Thus the low-energy $t_{2g}$
orbitals of the V($3d$) shell are split into an $a_{1g}$ and two degenerate
$e_g'$ orbitals. Formally the vanadium ion has the $3d^2$ valence configuration, i.e.
is in the V$^{3+}$ oxidation state. The $t_{2g}$ orbital degrees of freedom
appear to play a central role for the intriguing physics of this transition-metal 
oxide and Castellani {\sl et al.}~\cite{cas78} were the first to provide a detailed 
account of the complex correlated electronic structure in these local terms.

The finite-temperature phase diagram, taken from the original work of
McWhan and coworkers~\cite{mcw71,mcw73}, is
shown in Fig.~\ref{fig:v2o3}. In this work, V substitution by Ti (Cr)
implies positive (negative) pressure. It displays three major phases,
namely paramagnetic metallic (PMM), paramagnetic insulating (PMI) as
well as an antiferromagnetic insulating (AFI) regime. The transition
to the latter AFI phase at lower temperatures is also associated with
a structural transition to a monoclinic low-symmetry
structure.~\cite{ezh99} Furthermore, additional phase-diagram studies
in the vanadium-deficient V$_{2-y}$O$_3$ regime revealed the existence
of a metallic spin-density-wave (SDW) phase.~\cite{bao93}
Interestingly, that phase appears not to be a form of precursor to the
much more extended AFM ordering of the insulator. On the contrary it
seems that the magnetic short-range order within PMM and PMI is closer
to the SDW ordering. Therefore the magnetic ordering in the AFI phase
may be closely related to the structural change, even involving
additional orbital ordering.~\cite{bao97} Due to this additional
complexity in connection with the AFI phase we concentrate in this
work only on the phase equilibrium between PMM and PMI. A complete
description of the V$_2$O$_3$ phase diagram including the magnetically
ordered phases will be postponed to future studies.

\subsection{LDA characterization and local projections}
\begin{figure}[b]
\centering
\includegraphics*[width=7cm]{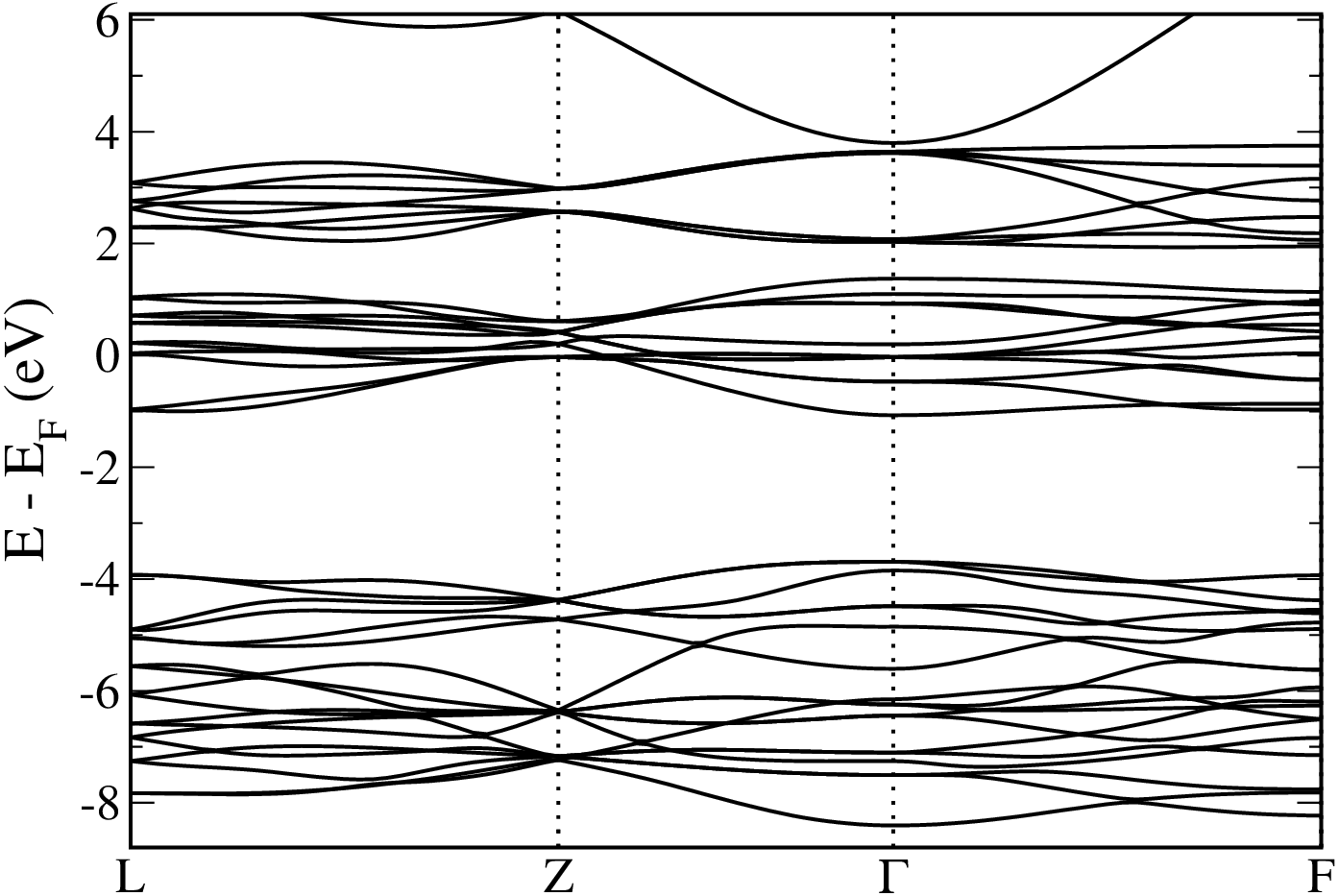}
\caption{LDA band structure of V$_2$O$_3$.
\label{fig:ldabands}}
\includegraphics*[width=7cm]{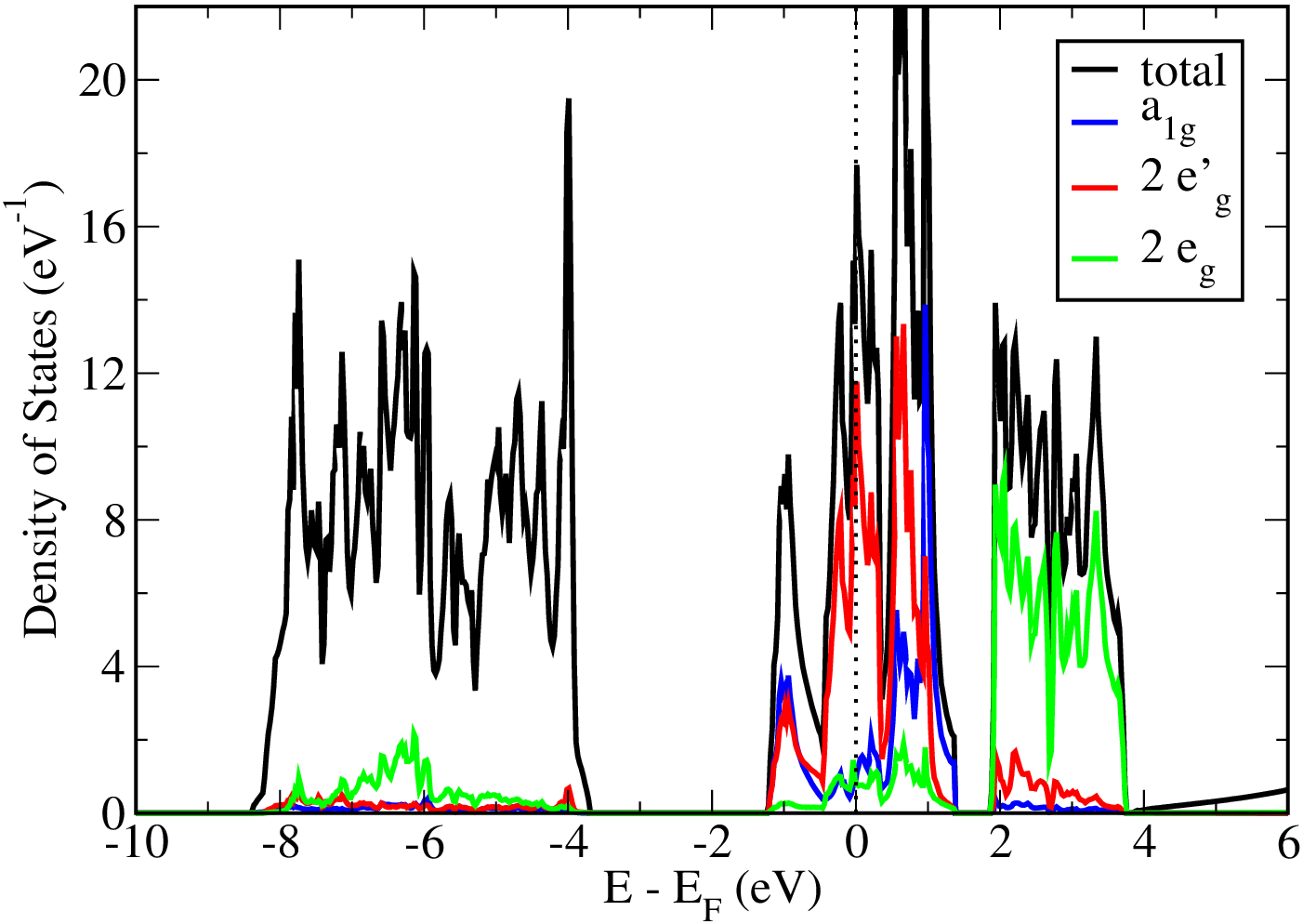}
\caption{(Color online) Total LDA density of states of V$_2$O$_3$ and local 
DOS within the symmetry-adapted V($3d$) basis with range $r_c$=2.0 a.u.
\label{fig:ldados}}
\includegraphics*[width=7cm]{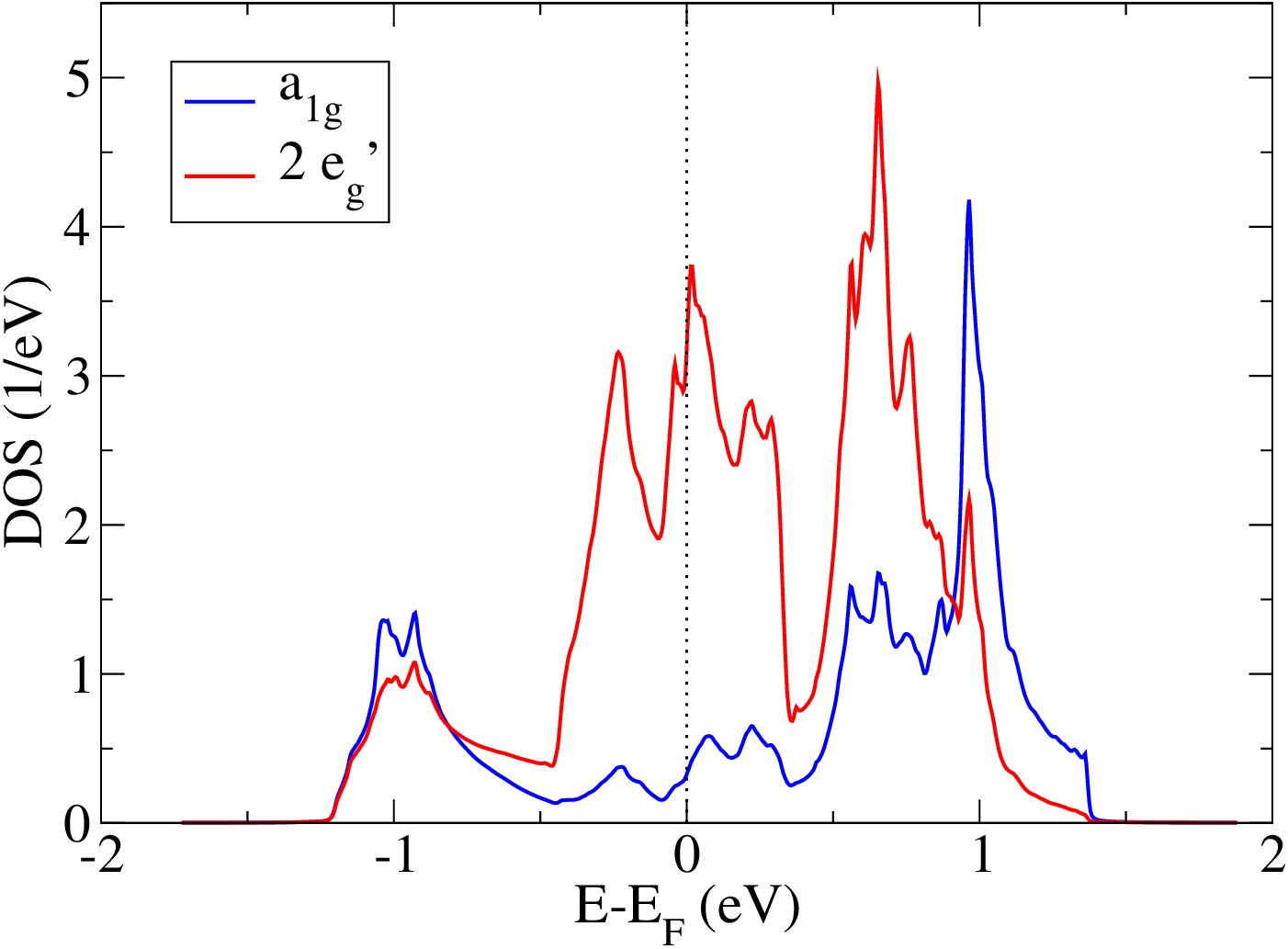}
\caption{(Color online) $a_{1g}$ and $e_g'$ LDA-DOS on the basis of the
orthonormalized projected local orbitals, whereby the range of the 
$\{\vert \chi_{\uR m}^{\hfill}\rangle\}$ was also $r_c$=2.0 a.u.}
\label{fig:projdos}
\end{figure}

A thorough first-principles DFT(LDA) description of metallic V$_2$O$_3$ at
normal pressure and without doping has initially been given by
Mattheiss.~\cite{mat94} Here we only summarize the most relevant features
as they evolve from our MBPP investigation. Fig.~\ref{fig:ldabands} shows
the LDA band structure along high-symmetry lines within the first Brillouin zone.
It is evident that the bands group in a way canonical for many transition-metal 
oxides. The larger block below the Fermi level $\epsilon_{\rm F}$ in the range 
$[-8,-4]$ eV is dominated by oxygen $2p$ orbital weight, while the unoccupied
block within $[2,4]$ eV stems majorly from vanadium $e_g$ orbitals. This encoding
is visualized in the density-of-states (DOS) plot of Fig.~\ref{fig:ldados} from
local projections onto the symmetry-adapted cubic-harmonic angular-momentum 
channel of the V$(3d)$ basis. The band manifold of
width $W$$\sim$2.6 eV around $\epsilon_{\rm F}$ is mostly composed of $a_{1g}$ and
$e_g'$ orbitals with only minor inter-mixing of V($e_g$) and O($2p$). Note that
especially the $a_{1g}$ character shows a prominent bonding-antibonding signature
in the DOS of this low-energy region.

From the LDA result the set of local orbitals 
$\{\vert \chi_{\uR m}^{\hfill}\rangle\}$ to be utilized in the local projections  
defined in eq.~(\ref{Eqn:PROJDEFINITION}) are here chosen to be given by the 
linear combinations of pseudized atomic $V(3d)$ functions that diagonalize the 
orbital density matrix on each of the four symmetry-equivalent vanadium ions within
the unit cell. This is often referred to as the crystal-field basis, whereby here
the $a_{1g}$ level is higher in energy than the $e_g'$ one. Figure~\ref{fig:projdos} 
exhibits the local DOS obtained from the
projections using for ${\cal W}$ the low-energy $t_{2g}$ manifold. In the following
we will only concentrate on these minimal projected local orbitals and will not
elaborate on the possible cases of larger energy windows ${\cal W}$, i.e. such
ones that also cover the high-energy occupied/unoccupied band manifolds. Note that
the local orbital DOS in Fig.~\ref{fig:ldados} and Fig.~\ref{fig:projdos} differ
on principle grounds (see also appendix~\ref{ldaucsc}). In 
the latter case it is computed in the basis of orthonormalized orbitals, whereas in
the former case it is calculated from projections onto angular-momentum channels without
proper final normalization (i.e. no radial orbital function involved). From the 
orthonormalized projected local orbitals one retrieves the occupations 
$n_{\rm LDA}(a_{1g})$=0.57 and $n_{\rm LDA}(e_g')$=1.43 
(summed over both $e_g'$ orbitals), respectively. 
These values differ by about 0.05 electrons towards stronger orbital polarization 
compared to the numbers presented in Ref.~\onlinecite{pot07} from a Wannier construction 
within the $N$th-order muffin-tin-orbital method.~\cite{and00}
\begin{figure}[t]
\centering
\includegraphics*[width=4.25cm]{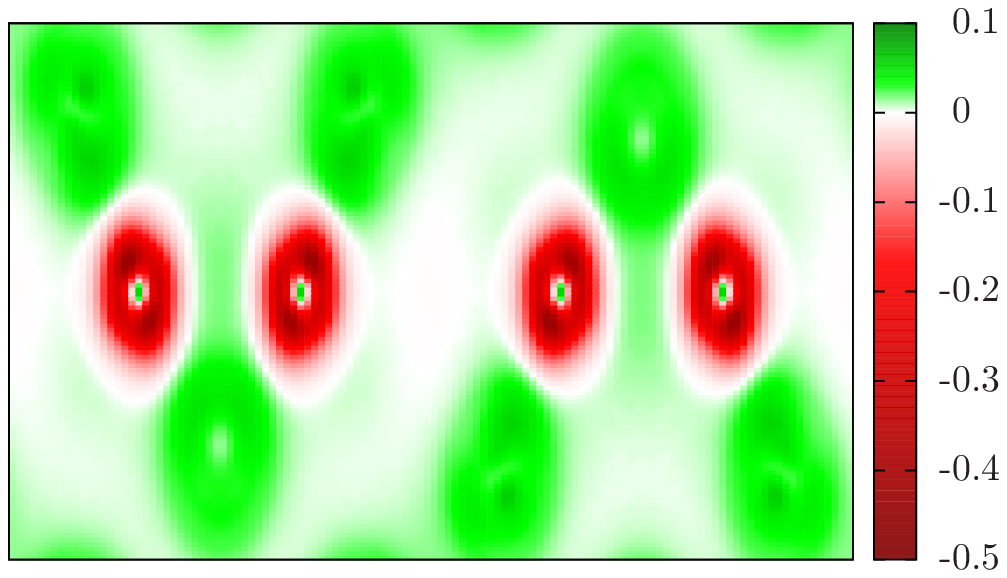}
\includegraphics*[width=4.25cm]{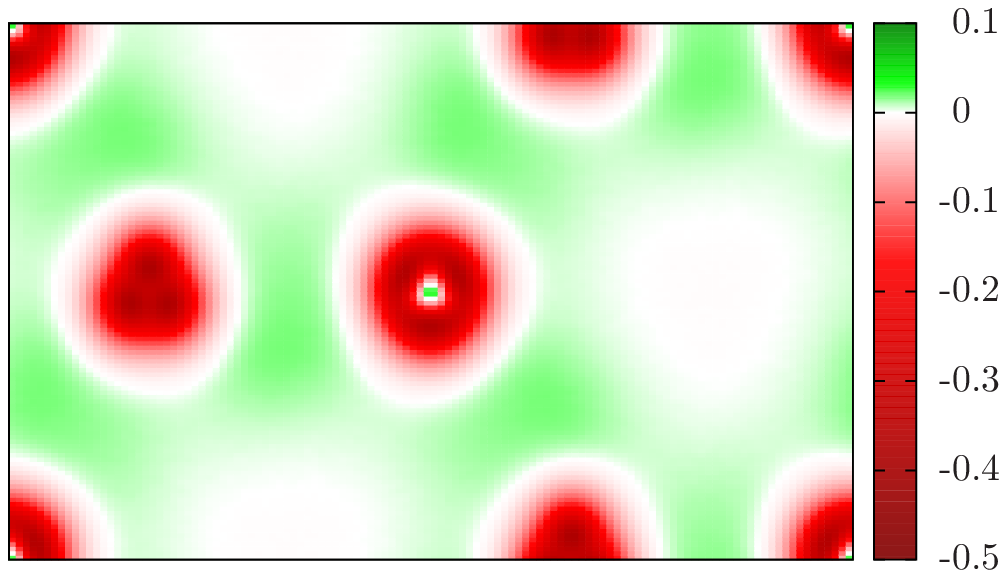}
\caption{(Color online) LDA bond charge densities
$\rho^{\rm crystal}(\ur)$$-$$\rho^{\rm atomic}(\ur)$ for V$_2$O$_3$ within the 
$xz$-plane (left) and the $xy$-plane (right). Note that in the latter case 
the two central V ions are not at the same height and therefore appear different.
\label{fig:lda_cd}}
\end{figure}

The stronger ionic character of the transition-metal oxide compared to ordinary
metals or intermetallic compounds becomes clear from the plot of the bonding
charge densities in Fig.~\ref{fig:lda_cd}. The latter function is defined as
the difference between the crystal valence charge density and the superposed atomic
valence charge densities. The charge transfer from vanadium to oxygen is obvious,
but also the expected charge accumulation in the interstitial region is visible.

\subsection{LDA+DMFT: finite-temperature phase equilibria}
The LDA-only description does not account for a metal-insulator
transition (MIT) in V$_2$O$_3$. We model the interacting problem on
the many-body level within a three-orbital ($a_{1g}$,~$2e_g'$)
multi-site (four V ions in the corundum unit cell) generalized Hubbard
model employing the complete rotational invariant Coulomb interactions
on the local level. For the parametrization of the Coulomb integrals
we choose the values $U$=5~eV and $J$=0.93~eV, as already utilized in
earlier simplified LDA+DMFT studies for V$_2$O$_3$.~\cite{hel01} The
following results are obtained from our MBPP-code interface of
LDA+DMFT. A second implementation within the PAW approach is also
briefly discussed in the appendix.

\subsubsection{Lattice expansion and temperature variation for fixed $c/a$ ratio}
\begin{figure}[b]
\centering
\includegraphics*[width=8.5cm]{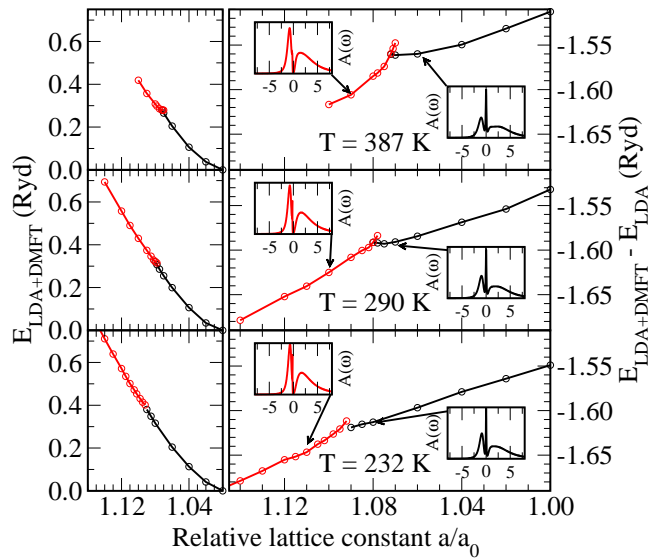}
\caption{(Color online) MIT with negative pressure, i.e. increase of the lattice
constant for various temperatures within LDA+DMFT. Left: total energy
$E_{\rm LDA+DMFT}$ normalized to the value at the equilibrium lattice constant, 
right: energies with respect to the LDA energy $E_{\rm LDA}$ for
each given lattice constant. 
\label{fig:mit-energy}}
\end{figure}

The MIT between the PMM and the PMI phase with negative pressure is depicted in 
Fig.~\ref{fig:mit-energy} for selected $T$. While the $p$$<$0 scenario is realized 
experimentally via Cr doping, it is here provided in a simple way by increasing the 
lattice constant $a$ starting from its experimental~\cite{der70} equilibrium value
$a_0$=4.95\AA. Notably we also first keep the $c/a$ ratio fixed to its value 
$c/a$=2.83 at ambient pressure and temperature. The effect of relaxing that ratio
will be discussed in section~\ref{relax-ca}. For the moment, this approximate theoretical 
approach proves to be sufficient to describe the key features of the V$_2$O$_3$ phase 
diagram above room temperature. 
\begin{figure}[t]
\centering
\includegraphics*[width=7.5cm]{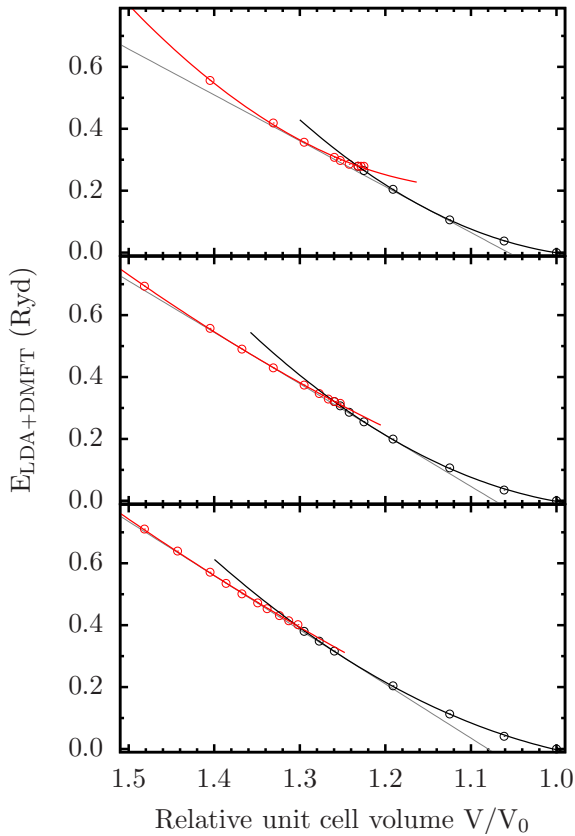}
\caption{(Color online) Tie-line construction for the first-order MIT with
negative $p$ for $T$=387 K, 290 K, 232 K (from top to bottom).
\label{fig:tieline}}
\end{figure}

One could connect this approach to the physical pressure $p$ in a simple way by 
defining $p$=$-\partial E/\partial V$. It is seen that the theoretical formalism 
reveals the pressure-induced first-order MIT with the correct positive sign of the 
slope $\partial T_{\rm MIT}/\partial p$ from experiment (compare 
Fig.~\ref{fig:v2o3}). However the changes that occur in the lattice constant
with $T$ along the phase boundary are non-surprinsingly larger (roughly by a factor 
5-10) than in experiment.~\cite{mcw69} The neglect of electronic and phononic entropy 
contributions (and presumably also non-local correlations) may be blamed.
Nevertheless the change of curvature of the respective total energies elucidates the 
expected softening of the lattice with increasing temperature from the decrease of 
the bulk modulus $B$$\sim$$\frac{\partial^2 E}{\partial V^2}$. Fig.~\ref{fig:tieline} 
displays the tie-line construction for the first-order transition between the metallic and 
insulating phase. The volume jump at the transition becomes obvious, albeit again
the retrieved pressure from the slope of the common tangent is about an order of 
magnitude too high compared to experiment. But note that the increased stability range
of the PMI phase with volume at larger $T$ is a direct result of the calculations. 
For $T_{\rm crit}$$\sim$400 K the line of first-order transitions at negative pressure 
exibits a solid-solid critical end point and a continuous path from the metallic to the 
insulating regime opens. One may already recognize that the shifted quasiparticle peak at 
the lower gap edge in the spectral function of the insulator has vanished for $T$=387 K,
which may signal the immediate strongly incoherent regime close to the critical end 
point.
\begin{figure}[t]
\centering
\includegraphics*[width=8cm]{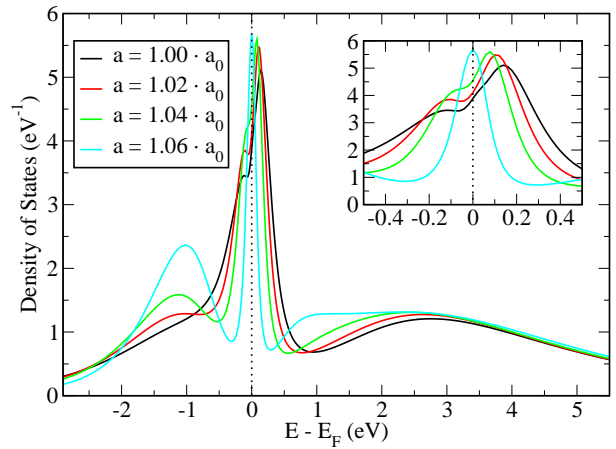}
\caption{(Color online) Evolution of the metallic spectral function with
negative pressure for $T$=387 K. The inset shows a blow-up of the features around
the Fermi level. 
\label{fig:corrstrength}}
\end{figure}

The pressure-dependent investigation describes moreover the increase of correlation
strength when approaching the critical $p$ deep from the metallic regime. This is
documented in Fig.~\ref{fig:corrstrength} where we plot the local spectral function
with increasing lattice constant. Strong transfer of spectral weight from the 
low-energy region to the high-energy Hubbard bands is observable and marks the
evolution towards the MIT with negative pressure at constant interaction strength.
Furthermore a shifting of the dominant quasiparticle peak towards the Fermi level 
with increasing the lattice constant may be recognized. Of course, the 
growing lattice distances also weakens the metallic screening and therefore should 
lead on simple grounds to an effective increase of the mutual Coulomb interaction 
between the electrons. Such an effect is here at least in simplest terms describable by 
the charge-self-consistent reaction to the applied lattice expansion. Note however in 
this context that recent photoemission studies point to a rather constant $U$ value
on the different sides of the phase-boundaries.~\cite{fuj11}
\begin{figure}[b]
\centering
\includegraphics*[width=8.5cm]{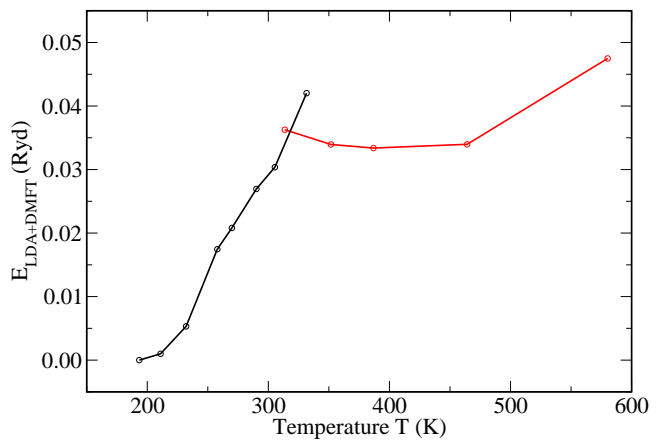}
\caption{(Color online) MIT with temperature for the fixed lattice constant
$a$=1.08$a_0$, i.e. at finite negative pressure.
\label{fig:mittemp}}
\end{figure}

In order to trace the MIT with temperature in some detail, we plot in 
Fig.~\ref{fig:mittemp} the intrinsically $T$-dependent LDA+DMFT energy now at fixed 
elongated lattice constant. Starting from the low-temperature metallic regime, 
$E_{\rm LDA+DMFT}(T)$ deviates from a simple functional behavior at $T$$\sim$270 K,
displaying an overall double-parabolic structure. The MIT takes place around 
$T_{\rm MIT}$$\sim$310 K in good accordance with the expected experimental region
of the temperature-induced PMM$-$PMI transition. Surprisingly, the energy
$E_{\rm LDA+DMFT}(T)$ of the PMI phase appears rather flat in the temperature regime
$[350,450]$ K, which is just in the neigborhood of the experimental critical end point.

\subsubsection{Charge densities and orbital resolution for fixed $c/a$ ratio}
\begin{figure}[b]
\centering
\includegraphics*[width=4.25cm]{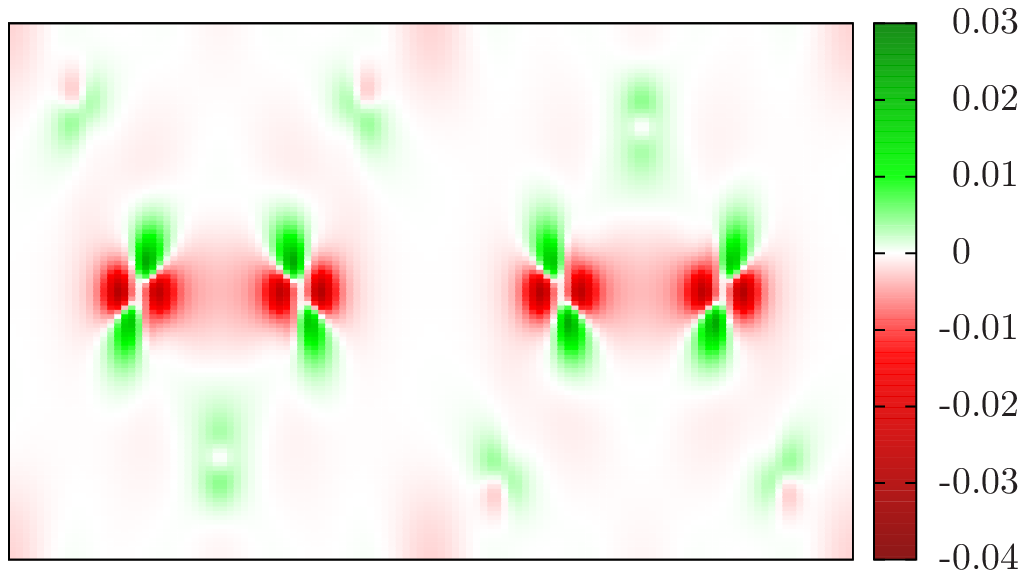}
\includegraphics*[width=4.25cm]{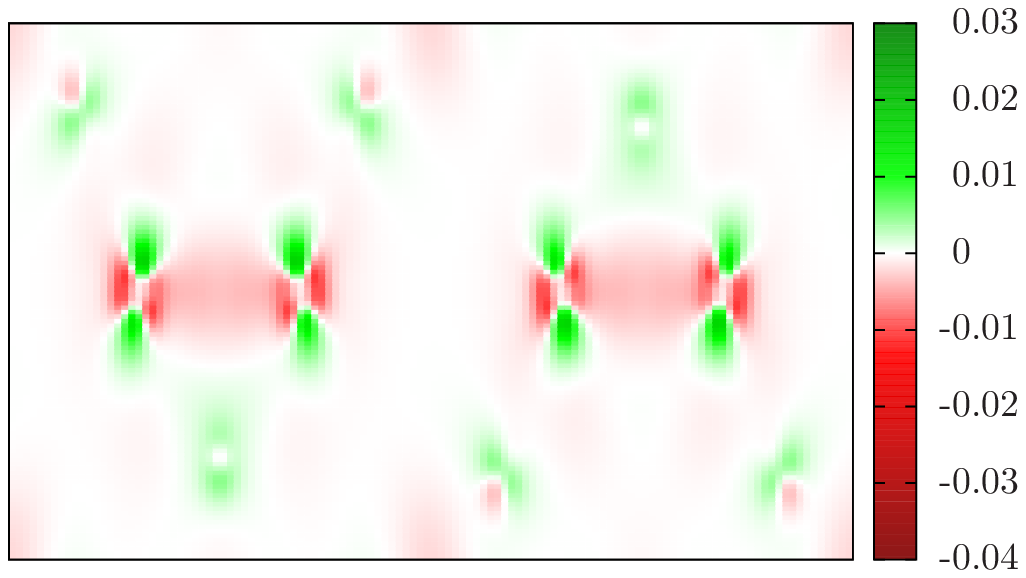}\\
\includegraphics*[width=4.25cm]{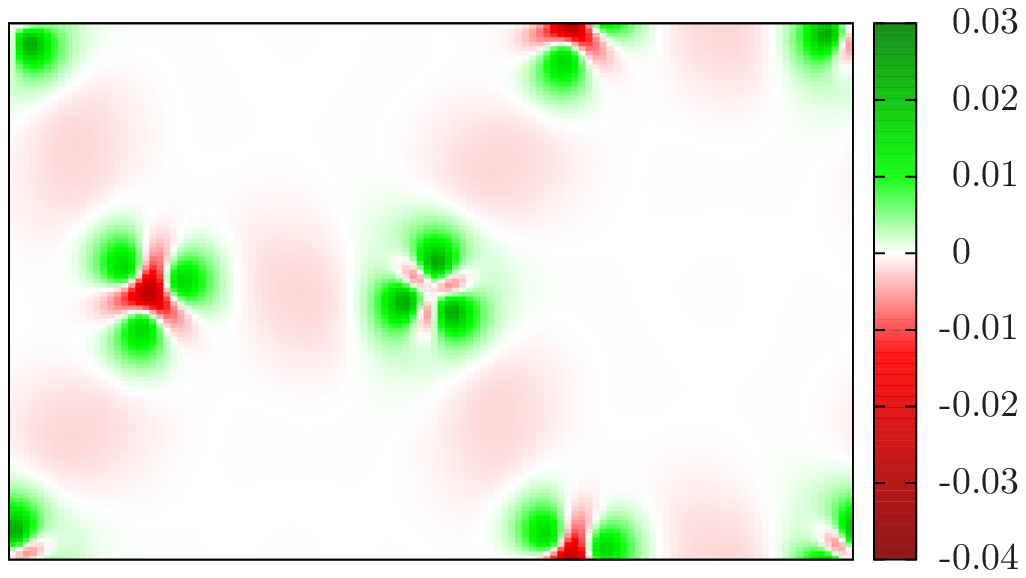}
\includegraphics*[width=4.25cm]{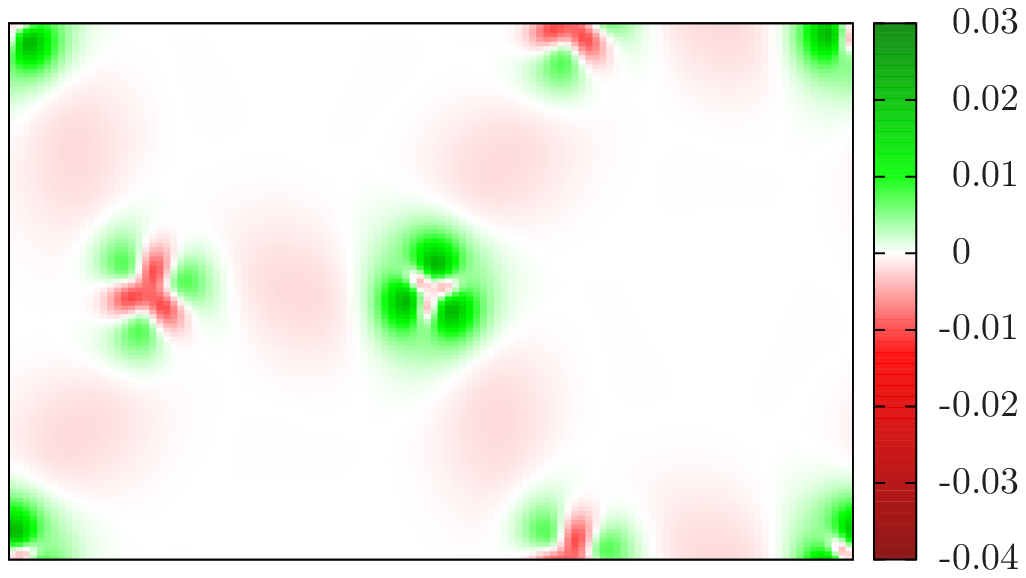}\\
\caption{(Color online) Differences of the electronic charge densities
$\rho_{\rm DMFT}(\ur)$$-$$\rho_{\rm LDA}(\ur)$ at $T$=387 K for the metal with
$a$=$a_0$ (left) and the insulator with $a$=1.1$a_0$ (right) within the 
$xz$-plane (top) and the $xy$-plane (bottom). 
\label{fig:chargedens2}}
\end{figure}
So far we concentrated on the integral impact of the electronic correlation on
the finite-temperature properties of the V$_2$O$_3$ system. However, for a deeper 
understanding of the underlying driving forces it is also important to shed 
light on the possibly distinct behavior of individual microscopic degrees of freedom 
and most notably on those of orbital kind. 
Concerning the distinct orbital occupations with temperature and negative pressure no
dramatic effects occur in the correlated electronic structure. In line with previous
postprocessing studies,~\cite{kel04,pot07} within charge-self-consistent LDA+DMFT
the $a_{1g}$ orbital filling is generally reduced compared to the LDA value
(and correspondingly the $e_g'$ filling is increased). For the equilibrium volume
and $T$=232 K the numbers write as $n_{\rm DMFT}(a_{1g})$=0.48 and 
$n_{\rm DMFT}(e_g')$=1.52. A real-space discrimination of
these orbital filling differences between LDA+DMFT and LDA on the basis of the 
respective charge densities is displayed in Fig.~\ref{fig:chargedens2} for $T$ 
close to the critical end point. In general, a localization effect takes place,
whereby charge from the interstitial region is transfered closer to the atomic sites.
Thereby the $a_{1g}$ orbital (pointing roughly along the $z$-axis) looses charge in 
the correlated electronic structure, whereas the $e_g'$ orbitals gain.~\cite{kel04,pot07} 
In the insulating regime less charge is transfered from the interstitial part, but
note that here the lattice constant is also larger.
\begin{figure}[t]
\centering
\includegraphics*[width=4.25cm]{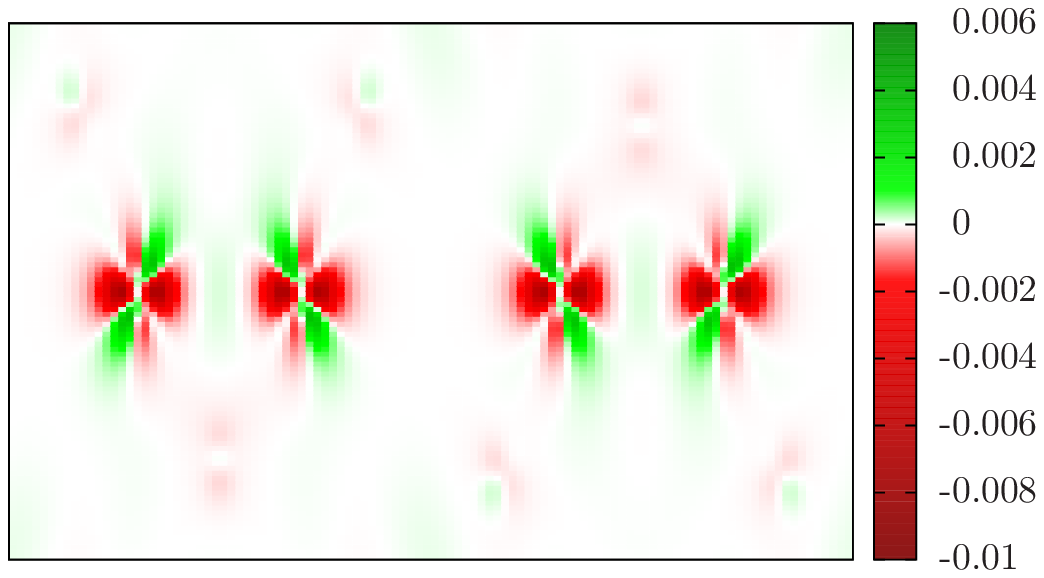}
\includegraphics*[width=4.25cm]{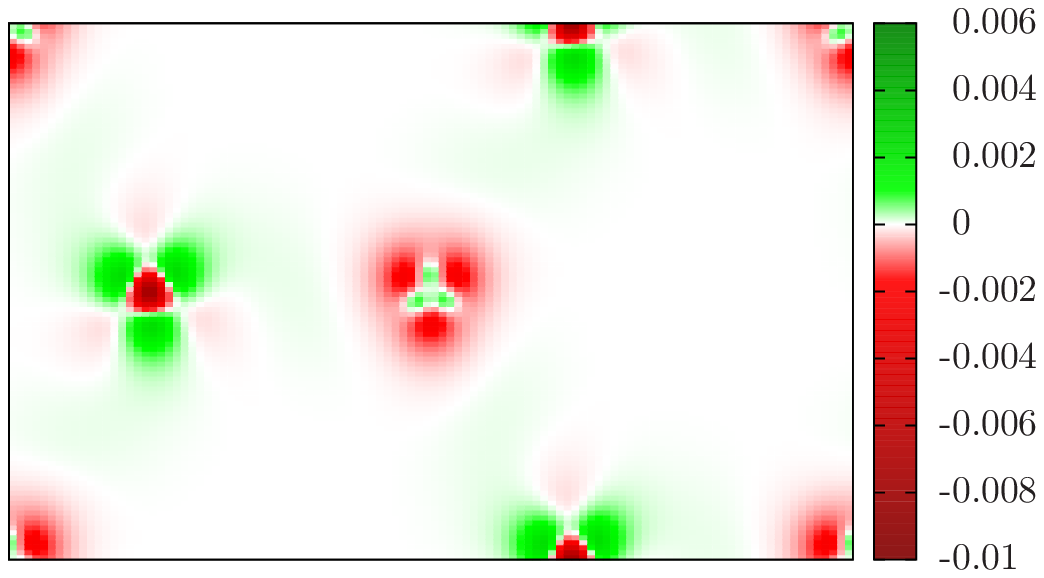}
\caption{(Color online) Difference 
$\rho_{\rm DMFT}^{\rm met}(\ur)$$-$$\rho_{\rm DMFT}^{\rm ins}(\ur)$
between the LDA+DMFT charge densities in the metallic ($T$=232 K) and in the 
insulating ($T$=387 K) regime for $a$=1.08$a_0$ within the $xz$-plane (left) 
and the $xy$-plane (right). 
\label{fig:chargedens1}}
\end{figure}
\begin{figure}[t]
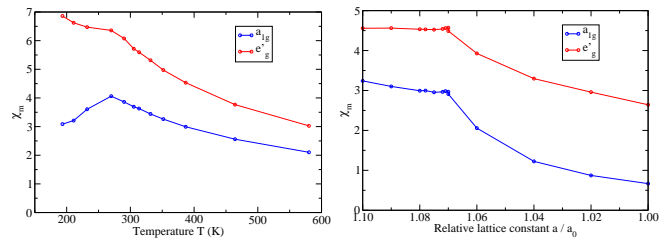

\centering
\includegraphics*[width=4.25cm]{chi_m_alat108.eps}
\includegraphics*[width=4.25cm]{chi_m_beta30.eps}\\
\caption{(Color online) Orbital-resolved local spin susceptibilities with
temperature (left) and negative pressure (right). 
\label{fig:chis}}
\end{figure}

Within the correlated scheme, increasing the negative $p$, i.e. enhancing the lattice 
constant, yields a slight filling increase of the $a_{1g}$ orbital. When raising $T$ for 
$a$=1.08$a_0$, so that the system shows a temperature-induced MIT,
the same trend occurs. The same effect with temperature was already theoretically observed
in Ref.~\onlinecite{pot07}. Thus the calculation reveals an increase (decrease) in the 
occupation (of the order of a few percent) for the $a_{1g}$ ($e_g'$) orbital in the 
insulator compared to the metal. It is again instructive to visualize directly the 
changes in the self-consistent correlated charge density. Figure~\ref{fig:chargedens1} 
depicts the differences in the LDA+DMFT charge density $\rho_{\rm DMFT}(\ur)$ at the 
different temperatures associated with the metallic and the insulating phase. It is 
seen that here the charge transfers are marginal, mainly showing the $a_{1g}$ orbitals
gaining some charge against the $e_g'$ with $T$. As an effect of the temperature 
raise, the interstitial inbetween the V ions appears to loose some small weight, in
the spirit of effective localization at high $T$.

Finally, we turn to a brief look on the magnetic response. The orbital-resolved 
local spin susceptibility $\chi$, plotted in Fig.~\ref{fig:chis}, shows for both 
orbital contributions the expected Curie-Weiss-like tail at higher temperatures, but for
fixed lattice constant a non-monotonic behavior below $T$$\sim$270 K emerges, i.e. in the
same range where the non-trivial characteristic in $E_{\rm LDA+DMFT}(T)$ was observed. The
quenching of $\chi_{a_{1g}}$ in that regime needs further study and might be 
interesting in the context of orbital-ordering in the low-temperature AFI phase.
As expected, for constant $T$ the increase of the lattice constant leads also to nearly
constant $\chi$ in the PMI phase.

\subsubsection{Effect of relaxing the $c/a$ ratio\label{relax-ca}}
\begin{figure}[t]
\centering
\includegraphics*[width=8cm]
{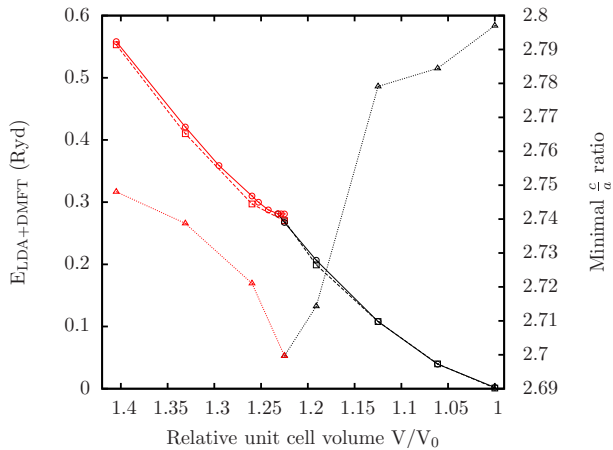}\\
\caption{(Color online) Left panel:
Total enery vs. volume for fixed $c/a$ (solid) and relaxed
$c/a$ (dashed). Right panel: Relaxed $c/a$ values vs. volume (dotted). Black lines mark the metallic
solution, red (grey) lines the insulating one. All data for $T$=387 K.
\label{fig:ca-relax}}
\end{figure}

From experiment it is known that the $c/a$ ratio is lowered when passing from the
metal to the insulator with negative pressure.~\cite{mcw69,rod10,rod11} In order to
account for that effect we relaxed $c/a$ for each volume $V$ within charge-self-consistent
LDA+DMFT at $T$=387 K by computing the total energy for selected $c/a$ values and
finding the minimum $E(V,c/a)$ via a polynomial fit to the data points. Note that
$c/a$ also varies substantially with temperature,~\cite{bald08} however as a proof of
principles we here only followed its evolution with expanding unit-cell volume. The 
results are shown in Fig.~\ref{fig:ca-relax}. No dramatic effect results in the global 
energetics, however the respective energy gain, especially in the PMI phase, is clearly 
visible. While in experiment the $c/a$ value varies within the interval 
$[2.78,2.88]$,~\cite{rod11} the given range $[2.70,2.80]$ is somewhat larger from the 
calculations but still within the right ballpark.
In accordance with the experimental data the PMM phase has a larger $c/a$ than
the PMI phase. Interestingly, the minimum ratio is just reached in the transition region
with respect to the volume, i.e. $c/a(V)$ becomes most soft close to the negative-pressure
driven MIT. Of course, at the latter first-order transition a jump in $c/a$ will take
place in line with the volume jump corresponding to the tie-line construction.

A relaxation of $c/a$ also affects the trigonal crystal-field splitting $\Delta_{\rm t}$ 
between $a_{1g}$ and $e_g'$ orbitals. Because of the lowering of $c/a$ with negative $p$
and hence a weakening of the distortion of the VO$_6$ octahedra a reduction of 
$\Delta_{\rm t}$ is expected. Indeed the calculations reveal a shrinking of $\Delta_t$ from
the PMM to the PMI phase. Note that thereby $\Delta_{\rm t}$ is determined from the sole
Hamiltonian part of the interacting problem stemming from the local-orbital projections 
acting on the LDA+DMFT converged Bloch states. A second contribution to the resulting
{\sl effective} crystal-field splitting $\Delta_{\rm eff}$ is given by the real part of the 
self-energy difference between $a_{1g}$ and $e_g'$ at zero frequency, i.e. 
$\Delta_{\rm eff}$=$\Delta_{\rm t}+\Re\Sigma_{a_{1g}}(0)-\Re\Sigma_{e_g'}(0)\,$.~\cite{pot07} However the latter difference differs only little between PMM and PMI and thus no
dramatic changes in the occupations occur. 

Contrary to that, former post-processing LDA+DMFT studies with an interaction-driven 
picturing of the difference between PMM and PMI phase resulted in rather strong 
orbital-polarized solutions for the insulating case.
That went along with an enhanced value for $\Delta_{\rm eff}$ due to large 
self-energy effects. Note that in the charge-self-consistent LDA+DMFT framework a clear-cut
separation into one-body and many-body contributions to $\Delta_{\rm eff}$ is not that
simple anymore. The reason is that during the self-consistency cyle the hoppings also change because
of the self-energy effects, contrary to post-processing LDA+DMFT. In our work we did not 
change the interaction parameters within the different phases, in line with recent 
experimental work.~\cite{fuj11} We however tried to change our global $(U,J)$ parameters 
in a certain range to look for the possibility of orbital-polarized solutions. Because of 
the complexity of the problem depending on the interplay
of hopping and many-body effects within an evoluting crystal-structure evolution 
temperature and pressure, we may not exclude a certain setup that allows for orbitally
polarized solutions. Yet within our studies we did not find clear evidence for this behavior.

\section{Discussion and Summary}
In this work the advancement of the LDA+DMFT methodology, namely the implementation
of a complete charge-self-consistent scheme with total-energy calculation, build on 
a pseudopotential band-structure code using plane waves and localized functions, 
was documented and the principle formalism of interfacing LDA and
DMFT utilizing projected local orbitals was reviewed. In the calculations we observed
that the charge-self-consistent framework, at least when using a minimal energy 
window for the projected local orbitals, leads for fixed values of the Hubbard $U$ to 
somewhat smaller electronic correlations than the elder post-processing scheme. Also
orbital polarizations are generally slightly weaker in the new complete methodology.
This outcome might be not that surprising since the now possible reaction of the
charge density to the self-energy effects may lead to additional screening effects.
Further more detailed investigations of the charge-self-consistent technique are
needed in this respect and shall be hereby stimulated. 

As a proof of principles, this approach renders it possible to describe the first-order
character of the MIT in the challenging V$_2$O$_3$ system induced by negative pressure 
and temperature in accordance with the experimental phase diagram. The methodology is
in the position to describe the PMM/PMI phase boundary in a qualitative correct manner.
Yet the quantitative 
agreement concerning structural data and pressure in the transition region is still not 
perfect. The neglect of electronic and vibrational entropy terms may be a probable cause, 
also since the absolute value of the critical negative pressure $|p|$$<$1GPa is rather 
small compared to other pressure-driven MITs.~\cite{kun08}

At first sight, a qualitative difference concerns the respective orbital 
($a_{1g}$, $e_g'$) fillings in the PMM and PMI phase. Whereas polarized x-ray absorption
measurements together with multiplet calculations point to an increased orbital 
polarization towards less filled $a_{1g}$ in the insulating regime,~\cite{par00} our 
calculations result in a tendency towards slight orbital balancing in the PMI phase. The
named stronger orbital polarization in the insulating regime is indeed verified in 
post-processing LDA+DMFT for fixed lattice structure, larger $U$ and constant 
$T$.~\cite{kel04,pot07} It results there from the increased effective crystal-field 
splitting due to the strong electronic correlations.
However it is important to note, that in the experimental part of the work by 
Park {\sl et al.}~\cite{par00} the negative pressure regime was realized by non-isovalent
Cr substitution for V. Recent work by Rodolakis {\sl et al.}~\cite{rod10,rod11} compared
that Cr doping-driven (and thus implictly negative-pressure driven) MIT with a true 
pressure-driven MIT. The latter scenario was realized by increasing pressure on 
insulating (V$_{0.92}$Cr$_{0.28}$)$_2$O$_3$. It was observed that with true applied 
pressure the orbital occupations hardly vary at the MIT, in good accordance with our 
results. Hence rather strong orbital polarization in the PMI phase appears to be bound to 
the doping-driven realization of the insulating phase. Further theoretical studies that
explicitly treat the chemical doping within an e.g. supercell approach, should clarify
this issue. The former post-processing LDA+DMFT treatments followed the route of mainly
interaction-driven MIT, while from recent hard x-ray photoemission spectroscopy it was
concluded that the Hubbard $U$ does not change through the MIT.~\cite{fuj11}

Of course when it comes to the question of orbital polarization, the competition between 
crystal-field effects and Coulomb correlations depends on the choice of the interaction 
parameters $(U,J)$~\cite{lecproc} and on the energy window used for the projection onto 
the correlated subspace. Especially concerning the latter, e.g. inclusion of the 
O$(2p)$-dominated bands (and/or extending the many-body part to a five-orbital sector) 
will change the notion of the $3d$ orbitals and their overall occupations significantly. 
More detailed studies along those lines, also by utilizing ab-initio computed 
Coulomb integrals, are surely necessary.

But despite this need the current work renders it clear that there is advancement in the 
LDA+DMFT framework that sharpens the tool for strongly correlated materials 
investigations of finite-temperature phase competitions on an equal footing with analyses 
of the involvment of local degrees of freedom.

\begin{acknowledgments}
We wish to thank L.~Boehnke, P.~E.~Bl\"ochl, I.~Leonov, A.~I.~Lichtenstein, 
L.~Pourovskii, S.~Schuwalow and C.~Walther for fruitful discussions. Financial support
by the DFG-FOR 1346 is gratefully acknowledged. Computations have been 
performed at the Regionales Rechenzentrum (RRZ) of the Universit\"at Hamburg 
as well as the Juropa Cluster of the J\"ulich Supercomputing Centre (JSC) under
project number hhh08.
\end{acknowledgments}

\appendix
\section{Representation of charge densities in the Kohn-Sham basis sets}
\label{App:CDDFTBASIS}

In this appendix we summarize how to compute the given charge density
from the matrix $\underline {\underline {\Delta N}}^{(\underline
  k)}$ (as in eq. (\ref{Eqn:RHOREPRESENTATION})) in two possible
Kohn-Sham basis sets.

\subsection{Mixed-Basis Pseudopotential (MBPP)}
The mixed-basis pseudopotential (MBPP) approach~\cite{lou79,mbpp_code} uses 
normconserving pseuodpotentials~\cite{van85} and a 
combined basis of plane waves and modified localized atomic functions 
$\phi_{\alpha l m}^{\underline k}(\underline r)$ for the representation of
the crystal wave functions, written as
\begin{equation}
\Psi_{\underline k \nu}(\underline r) = \frac 1 {\sqrt{\Omega_{\mathrm C}}}
\sum \limits_{\underline G} \Psi_{\underline G} ^{\underline k \nu}
\mathrm e^{i (\underline k + \underline G) \underline r} + \sum
\limits_{\alpha l m} \beta_{\alpha l m}^{\underline k \nu}
\phi_{\alpha l m}^{\underline k}(\underline r)\;,
\end{equation}
where $\Psi_{\underline G}$ and $\beta_{\alpha l m}$ are the respective expansion
coefficients for atom $\alpha$ in the unit cell and angular-momentum numbers $lm$. 
The correlated charge density therefore consists of three parts
$\rho(\underline r)$=$\rho^{(1)}(\underline r)$$+$$\rho^{(2)}(\underline
r)$$+$$\rho^{(3)}(\underline r)$, corresponding to a plane-wave term, a
mixed term and a localized-function term. With the abbrevation
\begin{equation}
N_{\nu \nu'}^{(\underline k)} := 
f(\tilde{\epsilon}_{\underline k \nu}) \delta_{\nu
 \nu'} + \Delta N_{\nu \nu'}^{(\underline k)}
\end{equation}
they can be written as follows
\begin{eqnarray}
\rho^{(1)}(\underline r) &=& \frac 1 {\Omega_{\mathrm C}} \sum
\limits_{\underline k \nu \nu'} N_{\nu \nu'}^{(\underline k)}
\sum \limits_{\underline G \underline G'} \left ( \Psi_{\underline G}
^{\underline k \nu'} \right )^\ast \Psi_{\underline G'} ^{\underline k \nu} \mathrm
e^{i (\underline G' - \underline G) \underline r}\;,\\
\rho^{(2)}(\underline r) &=& \frac 2 {\sqrt{\Omega_{\mathrm
      C}}} \sum \limits_{\underline k \nu \nu'} \Re \Bigl[
  N_{\nu \nu'}^{(\underline k)} \sum \limits_{\underline G}
  \Psi_{\underline G} ^{\underline k \nu} \mathrm e^{i (\underline k +
    \underline G) \underline r}\nonumber\\
&&\hspace*{2cm} \sum \limits_{\alpha l m} \left (
  \beta_{\alpha l m}^{\underline k \nu'} \right )^\ast \left (
  \phi_{\alpha l m}^{\underline k}(\underline r) \right )^\ast \Bigr]\;,\\
\rho^{(3)}(\underline r) &=& \sum \limits_{\underline k \nu \nu'}
N_{\nu \nu'}^{(\underline k)} \sum \limits_{\alpha l m} \left
( \beta_{\alpha l m}^{\underline k \nu'} \right )^\ast \left (
\phi_{\alpha l m}^{\underline k}(\underline r) \right )^\ast\nonumber\\
&&\hspace*{2cm} \sum\limits_{\alpha' l' m'} \beta_{\alpha' l' m'}^{\underline k \nu}
\phi_{\alpha' l' m'}^{\underline k}(\underline r)\;.
\end{eqnarray}
The first term can be evaluated directly by Fourier transformation of
both wave functions to real space. The second and third term, which
are zero in the interstitial region (where the localized function have
decayed), are calculated in a straightforward way in an atom-centered
basis.

\subsection{Projector-Augmented Wave (PAW)}
The implementation within the PAW formalism is in line with Ref.~\onlinecite{ama12}. 
As shown in Ref.~\onlinecite{blo03}, charge densities here break down also into three 
parts, namely a plane-wave part $\tilde \rho(\underline r)$ and a
one-centre term from partial waves $\rho_{\underline R}^1 (\underline
r)$ and from pseudo-partial waves $\tilde \rho_{\underline R}^1
(\underline r)$ per atom at site $\underline R$ (omitting core
densities, which are not affected by our LDA+DMFT approach):
\begin{equation}
\rho(\underline r) = \tilde \rho(\underline r) + \sum
\limits_{\underline R} \left (\rho_{\underline R}^1 (\underline r) -
\tilde \rho_{\underline R}^1 (\underline r) \right )
\end{equation}
The plane-wave part $\tilde \rho(\underline r)$ can be calculated
directly as in eq. (\ref{Eqn:RHOREPRESENTATION}) from the PAW
pseudo wave functions $\vert \tilde \Psi_{\underline k \nu}
\rangle$. For the one-centre terms, the following one-centre density matrix may
be defined via
\begin{equation}
\mathcal D_{ij} = \sum \limits_{\underline k\nu\nu'} 
\langle \tilde p_i \vert \tilde \Psi_{\underline k \nu} \rangle
\left(f(\tilde{\epsilon}_{\underline k \nu}) \delta_{\nu \nu'}
+ \Delta N_{\nu \nu'}^{(\underline k)} \right )
\langle \tilde \Psi_{\underline k \nu'} \vert \tilde p_j \rangle\;.
\end{equation}
Here $\vert \tilde p_i \rangle$ are the projector functions from the
PAW formalism. With this definition, $\rho_{\underline R}^1
(\underline r)$ and $\tilde \rho_{\underline R}^1 (\underline r)$ can
be calculated as usual from the partial waves $\phi_i(\underline r)$
and from the pseudo partial waves $\tilde \phi_i(\underline r)$, i.e.
\begin{eqnarray}
\rho_{\underline R}^1 (\underline r) &=& \sum \limits_{i,j \in
  \underline R} \mathcal D_{ij} \phi_j^\ast (\underline r)
\phi_i(\underline r)\;,\\
\tilde \rho_{\underline R}^1 (\underline r) &=& \sum \limits_{i,j \in
  \underline R} \mathcal D_{ij} \tilde \phi_j^\ast (\underline r)
\tilde \phi_i(\underline r)\;.
\end{eqnarray}

\subsection{Comparison of the present charge-self-consitent scheme to standard 
LDA+U implementations\label{ldaucsc}}
\begin{figure}[t]
\centering
\includegraphics*[width=0.9\linewidth]{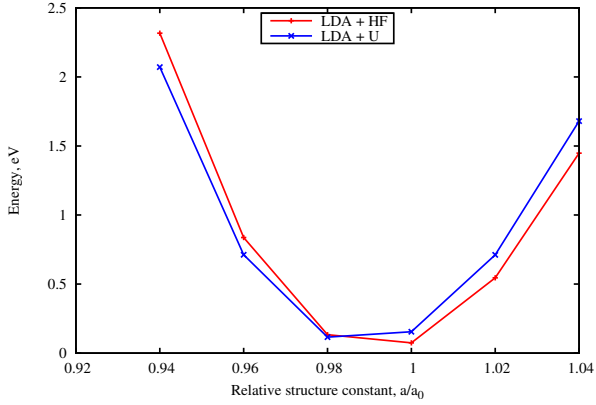}
\caption{(Color online) The equation of state of V$_2$O$_3$ calculated with 
the conventional LDA+U scheme as well as the charge-self-consistent scheme using
an Hartree-Fock solver for the DMFT impurity problem. 
\label{fig:hf_ldau}}
\end{figure}

In many implementations of the LDA+U scheme, originally designed for long-range 
ordered strongly correlated insulators, the local problem
is constructed by projecting onto a set of angular-momentum channels (i.e., spherical
or cubic harmonics) within a given range around the correlated site. For instance,
the LDA+U implementation in the MBPP code as well as in the Vienna Ab-initio 
Simulation Package (VASP)~\cite{kresse96,bengone00} is performed in such a way.
Here we want to show that the {\sl overall interfacing structure} of the present 
charge-self-consistent scheme of extending LDA gives similar results as the 
traditional LDA+U scheme, if a simple purely static mean-field approximation to
the local interacting problem is used.

As a simple test, one can compare the variation of the total energy
resulting from a conventional LDA+U calculation with the results obtained from
the present LDA+DMFT charge-self-consistent calculation now using the Hartree-Fock 
(HF) approximation for the DMFT impurity solver. 

We have calculated the equation of state for V$_{2}$O$_{3}$ within
the two approaches and the comparison is shown in Fig.~\ref{fig:hf_ldau}.
The LDA+DMFT(HF) (or, simply LDA+HF) calculation is performed using
the projected local orbitals as defined in eq.~(\ref{Eqn:PROJDEFINITION})
implemented in the VASP code, while the results for the LDA+U scheme are obtained 
using the standard VASP implementation.~\cite{bengone00} The results are rather
similiar and most of the differences may be due to the alternative choices for
the local projections. Note however that especially in the general LDA+DMFT context 
the local projections as defined in eq.~(\ref{Eqn:PROJDEFINITION}) are clearly 
superior (e.g. via the resulting well-defined local Green's function) to the simple 
angular-momentum-channel projections. 

\bibliography{bibextra}

\end{document}